\documentclass[aps]{revtex4}
\usepackage{amsfonts}
\begin{document}
\title{Superresonance effect from a rotating acoustic black hole and Lorentz symmetry breaking}
\author{M. A. Anacleto, F. A. Brito and E. Passos}
\email{anacleto, fabrito, passos@df.ufcg.edu.br}
\affiliation{Departamento de F\'{\i}sica, Universidade Federal de Campina Grande, Caixa Postal 10071, 58109-970 Campina Grande, Para\'{\i}ba, Brazil}
\begin{abstract} 
We investigate the possibility of the acoustic superresonance phenomenon (analog to the superradiance in black hole physics), i.e., the amplification of a sound wave by reflection from the ergoregion of a rotating acoustic black hole with Lorentz symmetry breaking. 
For rotating black holes the effect of superradiance corresponds to the situation where the incident waves has reflection coefficient greater than one, and energy is extracted from them. For an acoustic Kerr-like black hole its rate of loss of mass is affected by  the Lorentz symmetry breaking. We also have shown that for suitable values of the Lorentz violating parameter a wider spectrum of particle wave function can be scattered with increased amplitude by the acoustic black hole.

\end{abstract}
\maketitle
\pretolerance10000

\section{Introduction}

Acoustic black holes possess many of the fundamental properties of the black holes of general relativity and have been extensively studied in the 
literature \cite{matt,Volovik,Unruh}. The connection between black hole physics and the theory of supersonic acoustic flow was established in 1981 by Unruh \cite{Unruh} and has been developed to investigate the Hawking radiation 
and other phenomena for understanding quantum gravity. Hawking radiation is an important quantum effect of black hole physics. In 1974, Hawking combining Einstein's General Relativity and
Quantum Mechanics announced that classically a black hole does not radiate, but when we consider quantum effects emits thermal radiation at a temperature proportional to the horizon surface gravity.

Since the Hawking radiation showed by Unruh \cite{Unruh} is a purely
kinematic effect of quantum field theory, we can study the Hawking radiation process in completely
different physical systems. For example, acoustic horizons are regions where a moving fluid exceeds the local sound speed through 
a spherical surface and possesses many of the properties associated with
the event horizons of general relativity. In particular, the acoustic Hawking radiation when quantized appears as a flux of thermal phonons
emitted from the horizon at a temperature proportional to its surface gravity. Many fluid systems
have been investigated on a variety of analog models of acoustic black holes, including gravity wave~\cite{RS}, water
\cite{Mathis}, slow light \cite{UL}, optical fiber \cite{Philbin} and  electromagnetic waveguide \cite{RSch}. The models of superfluid helium II \cite{Novello}, atomic Bose-Einstein condensates \cite{Garay,OL} and one-dimensional Fermi degenerate
noninteracting gas \cite{SG} have been proposed to create an acoustic black hole geometry in
the laboratory.

The purpose of this paper is to study the superresonance phenomenon  considering the idea of the Lorentz symmetry breaking theories suggested in the seminal paper in Superstring Theory \cite{Kost} and further developed in Quantum Field Theory and General Relativity \cite{Colladay:1998fq, Kostelecky:2003fs, CFJ, JAK-PI} to investigate the relativistic version of acoustic black holes from the Abelian Higgs model \cite{Xian} with Lorentz symmetry breaking. A relativistic version of acoustic black holes has been presented in \cite{Xian,ABP} (see also \cite{Bilic}). This is also motivated by the fact that in high energy physics both strong Lorentz symmetry violation and quark gluon plasma (QGP) may take place together. Thus, it seems to be natural to look for acoustic black holes in a QGP fluid with Lorentz symmetry breaking  in this regime. Acoustic phenomena in QGP matter can be seen in Ref.~\cite{shk} and acoustic black holes in a plasma fluid can be found in Ref.~\cite{BH-plasma}.

Differently of the most cases studied, we consider the acoustic black hole metrics obtained from a relativistic fluid plus a term that violates the Lorentz symmetry. 
 The effects of this
set up is such that the fluctuations of the fluids are also affected. The sound waves inherit the broken Lorentz symmetry of the fluid, lose the Lorentz boost invariance and develops a birefringence phenomenon. As consequence the Hawking temperature is directly affected by the Lorentz-violating 
term. Analogously to Lorentz-violating gravitational black holes \cite{syb,adam}, the effective Hawking temperature of the acoustic black holes 
now is {\it not} universal for all species of particles. It
depends on the maximal attainable velocity of this species.
 Furthermore, the acoustic black hole metric can be identified with 
an acoustic Kerr-like black hole. The Lorentz violating term affects the rate of loss of mass of the black hole. We also have shown that for suitable values of
the Lorentz violating parameter a wider spectrum of particle wave function can be scattered with increased amplitude by the acoustic black hole. This
increases the superressonance phenomenon previously studied in \cite{Basak:2002aw,SBP}.

The paper is organized as follows. In Sec.~\ref{II} we apply the black hole metrics obtained in the extended Abelian Higgs model with the Lorentz-violating term, first introduced in \cite{ABP}. In Sec.~\ref{PV} we address the issue of perfect vortex. We find that even in Lorentz symmetry breaking fluids there is $no$ superresonance phenomenon, though the angular velocity is changed. In Sec.~\ref{conclu} we make our final conclusions.

\section{The Lorentz-Violating Model}
\label{II}
In this section we shall apply the acoustic black hole metrics obtained in the extension of the Abelian Higgs model with a modified scalar sector via scalar Lorentz-violating term \cite{ABP,Bazeia:2005tb}.
The Lagrangian of the Lorentz-violating Abelian Higgs model is
\begin{eqnarray}
\label{acao}
{\cal L}&=&-\frac{1}{4}F_{\mu\nu}F^{\mu\nu} +|D_{\mu}\phi|^2+ m^2|\phi|^2-b|\phi|^4+ k^{\mu\nu}D_{\mu}\phi^{\ast}D_{\nu}\phi,
\end{eqnarray}
where $F_{\mu\nu}=\partial_{\mu}A_{\nu}-\partial_{\nu}A_{\mu}$, $D_{\mu}\phi=\partial_{\mu}\phi - ieA_{\mu}\phi$ and $k^{\mu\nu}$  is a constant symmetric tensor implementing the Lorentz symmetry breaking.  The upper bound for its components are $k_{00}\leq 3.6\times10^{-8}$ \cite{codata} and ${\rm tr}(k_{ij})\leq3\times10^{-6}$ in relativistic and non-relativistic BEC theory \cite{Casana:2011bv}. In our present study, for the sake of simplicity,  we reduce the ten components of the tensor to two independent components by choosing  the following entries $k_{ii}=k_{00}\equiv\beta$ and $k_{0i}=k_{ij}\equiv\alpha$. While this is non trivial enough for our analysis other effects could also be achieved  by making other choices. We further assume all the components with magnitude around the shorter bound $k_{00}\leq 3.6\times10^{-8}$ . The tensor is now given by the form
\begin{equation}
k_{\mu\nu}=\left[\begin{array}{clcl}
\beta &\alpha &\alpha & \alpha\\
\alpha &\beta &\alpha &\alpha \\
\alpha &\alpha &\beta &\alpha\\
\alpha &\alpha &\alpha &\beta
\end{array}\right], \quad(\mu,\nu=0,1,2,3),
\end{equation}
being $\alpha$ and $\beta$ real parameters with magnitude around $k_{00}$. In a previous study \cite{ABP} following this theory we have found three- and two-dimensional acoustic metric describing acoustic black holes. In the following we shall focus on the planar acoustic black hole metrics to address the issue of superresonance phenomenon \cite{Basak:2002aw,SBP}.

\pretolerance10000

\subsection{The case $\beta\neq0$ and $\alpha=0$}
The acoustic line element in polar coordinates on the plane with Lorentz symmetry breaking, up to an irrelevant position-independent factor, in the `non-relativistic' limit ($v^2\ll c^2$) is given by \cite{ABP}
\begin{eqnarray}
ds^2=-\frac{(c^2-{\tilde\beta}_{-}v^2)}{{\tilde\beta}_{+}}dt^2-2(v_{r}dr+v_{\phi}rd\phi)dt+\frac{{\tilde\beta}_{+}}{{\tilde\beta}_{-}}(dr^2+r^2d\phi^2),
\end{eqnarray}
 where $\tilde{\beta}_{\pm}\equiv 1\pm\beta$, $c$ is the sound velocity in the fluid and $v$ is the fluid flow velocity.
We consider the flow with the velocity potential $\psi(r,\phi) = A\ln{r} + B\phi$  whose velocity profile in polar coordinates on the plane  is  given by \cite{matt}
\begin{eqnarray}
\vec{v}=\frac{A}{r}\hat{r}+\frac{B}{r}\hat{\phi}.
\end{eqnarray}
The transformations of the time and the azimuthal angle coordinates
\begin{eqnarray}
d\tau&=&dt+\frac{{\tilde\beta}_{+}Ardr}{(c^2r^2-{\tilde\beta}_{-}A^2)},
\\
d\varphi&=&d\phi+\frac{B{\tilde\beta}_{-}Adr}{r(c^2r^2-{\tilde\beta}_{-}A^2)}.
\end{eqnarray}
In the new coordinates, the metric becomes
\begin{eqnarray}
\label{ELB}
ds^2\!=\!\frac{{\tilde\beta}_{+}}{{\tilde\beta}_{-}}
\left[-\frac{{\tilde\beta}_{-}}{{\tilde\beta}_{+}^2}\left(1-\frac{{\tilde\beta}_{-}(A^2+B^2)}{c^2r^2}\right)d\tau^2
+\left(1-\frac{{\tilde\beta}_{-}A^2}{c^2r^2}\right)^{-1}dr^2
-\frac{2{\tilde\beta}_{-}B}{{\tilde\beta}_{+}c}d\varphi d\tau+r^2d\varphi^2\right].
\end{eqnarray}
The radius of the ergosphere is given by $g_{00}(r_{e}) = 0$, whereas the horizon is given by the coordinate singularity $g_{rr}(r_{h}) = 0$ \cite{matt},  that is 
\begin{eqnarray}
r_{e}=\frac{{\tilde\beta}_{-}^{1/2}}{c}\sqrt{{A^2+B^2}}, \quad r_{h}=\frac{{\tilde\beta}_{-}^{1/2}|A|}{c}.
\end{eqnarray}
We can observe from the equation (\ref{ELB}) that for $A > 0$ we are dealing
with a past event horizon, i.e., acoustic white hole and
for $A < 0$ we are dealing with a future acoustic horizon, i.e.,
acoustic black hole.
The metric is
\begin{eqnarray}
g_{\mu\nu}=\frac{{\tilde\beta}_{+}}{{\tilde\beta}_{-}}\left[\begin{array}{clcl}
-\frac{{\tilde\beta}_{-}}{{\tilde\beta}_{+}^2}\left[1-\frac{r_{e}^2}{r^2}\right] &\quad\quad 0& -\frac{{\tilde\beta}_{-}}{{\tilde\beta}_{+}}\frac{B}{cr}\\
0 & \left(1-\frac{r_{h}^2}{r^2} \right)^{-1}& 0\\
-\frac{{\tilde\beta}_{-}}{{\tilde\beta}_{+}}\frac{B}{cr} &\quad\quad 0 & 1
\end{array}\right],
\end{eqnarray}
and inverse $g^{\mu\nu}$
\begin{eqnarray}
\label{metrinv}
g^{\mu\nu}=\frac{{\tilde\beta}_{+}}{{\tilde\beta}_{-}}\left[\begin{array}{clcl}
-\frac{\tilde{\beta}_{+}^2}{\tilde{\beta}_{-}}\Gamma(r) &\quad\quad 0& -\frac{\tilde{\beta}_{+}B}{cr}\Gamma(r)\\
0 & \left(1-\frac{r_{h}^2}{r^2} \right)& 0\\
-\frac{\tilde{\beta}_{+}B}{cr}\Gamma(r) &\quad\quad 0 & \left[1-\frac{r_{e}^2}{r^2}\right]\Gamma(r)
\end{array}\right],
\end{eqnarray}
where $\Gamma(r)=\left[1-\frac{r_{h}^2}{r^2}\right]^{-1}$. 

We consider the Klein-Gordon equation for a linear acoustic disturbance $\psi(t,r,\phi)$ in the background metric (\ref{metrinv}), i.e.,
\begin{eqnarray}
\frac{1}{\sqrt{-g}}\partial_{\mu}(\sqrt{-g}g^{\mu\nu}\partial_{\nu})\psi=0.
\end{eqnarray}
We can make a separation of variables into the equation above as follows
\begin{eqnarray}
\psi(t,r,\phi)=R(r)e^{i(\omega t-m\phi)}.
\end{eqnarray}
The radial function $R(r)$ satisfies the linear second order differential equation, which turns out to be a simpler one dimensional Schroedinger problem
\begin{eqnarray}
\label{EQKG}
&&\left[\frac{\tilde{\beta}_{+}^2}{\tilde{\beta}_{-}}\omega^2-\frac{2\tilde{\beta}_{+}Bm\omega}{cr^2}-\frac{m^2}{r^2}\left(1-\frac{r_{e}^2}{r^2}\right)\right]R(r)
+\frac{1}{r}\left(1-\frac{r_{h}^2}{r^2}\right)\frac{d}{dr}\left[r\left(1-\frac{r_{h}^2}{r^2}\right)
\frac{d}{dr}\right]R(r)=0.
\end{eqnarray}
We now introduce the tortoise coordinate $r^{\ast}$ by using the following equation
\begin{eqnarray}
\frac{d}{dr^{\ast}}=\left(1-\frac{r_{h}^2}{r^2}\right)\frac{d}{dr}=\left(1-\frac{{\tilde\beta}_{-}A^2}{c^2r^2}\right)\frac{d}{dr},
\end{eqnarray}
which gives the solution
\begin{eqnarray}
r^{\ast}=r+\frac{\sqrt{{\tilde\beta}_{-}}|A|}{2c}\log{\left(\frac{r-\frac{\sqrt{{\tilde\beta}_{-}}|A|}{c}}{r+\frac{\sqrt{{\tilde\beta}_{-}}|A|}{c}}\right)}.
\end{eqnarray}
Observe that in this new coordinate the horizon $r_{h}=\frac{{\tilde\beta}_{-}^{1/2}|A|}{c}$
maps to $r^{\ast}\rightarrow-\infty$ 
while $r\rightarrow\infty$ corresponds to $r^{\ast}\rightarrow+\infty$.
Now, we consider a new radial function, $G(r^{\ast})=r^{1/2}R(r)$ and the modified radial equation obtained from (\ref{EQKG}),
in the asymptotic region ($r^{\ast}\rightarrow\infty$), that can be approximately written as
\begin{eqnarray}
\label{EG}
\frac{d^2G(r^{\ast})}{dr^{\ast2}}+\tilde{\omega}^2G(r^{\ast})=0, \quad \tilde{\omega}^2=\frac{\tilde{\beta}_{+}^2}{\tilde{\beta}_{-}}\omega^2.
\end{eqnarray}
We find for the equation (\ref{EG}) the simple solution
\begin{eqnarray}
\label{sl}
G(r^{\ast})=e^{i\tilde{\omega}r^{\ast}}+{\cal R}e^{-i\tilde{\omega}r^{\ast}}\equiv G_{A}(r^{\ast}).
\end{eqnarray}
Notice that the first term in equation (\ref{sl}) corresponds to ingoing wave and the second term to the reflected wave, so that ${\cal R}$ is the reflection coefficient as in usual studies of potential scattering.
The Wronskian of the solutions (\ref{sl}) can be computed to give
\begin{eqnarray}
{\cal W}(+\infty)=-2i\tilde{\omega}(1-|{\cal R}|^2).
\end{eqnarray}

Now, near the horizon region ($r^{\ast}\rightarrow-\infty$), we have
\begin{eqnarray}
\frac{d^2G(r^{\ast})}{dr^{\ast2}}+\left(\tilde{\omega}-m\tilde{\Omega}_{H}\right)^2G(r^{\ast})=0,
\end{eqnarray}
where, $\tilde{\Omega}_{H}=\Omega_{H}/\sqrt{\tilde{\beta}_{-}}$ and $\Omega_{H}=Bc/A^2$ is the angular velocity of the acoustic black hole. We suppose that just the solution identified by ingoing wave is physical, so that
\begin{eqnarray}
G(r^{\ast})={\cal T}e^{i\left(\tilde{\omega}-m\tilde{\Omega}_{H}\right)r^{\ast}}\equiv G_{H}(r^{\ast}).
\end{eqnarray}
The undetermined coefficient ${\cal T}$ is the transmission coefficient of our one dimensional Schroedinger problem. Now the Wronskian of the solution is
\begin{eqnarray}
{\cal W}(-\infty)=-2i\left(\tilde{\omega}-m\tilde{\Omega}_{H}\right)|{\cal T}|^2
\end{eqnarray}
Because both equations are approximate solutions of the asymptotic limit of the modified radial equation, the Wronskian is constant and then
${\cal W}(+\infty)={\cal W}(-\infty)$. 
Thus, we obtain the reflection coefficient
\begin{eqnarray}
\label{refle}
|{\cal R}|^2=1-\left(\frac{\tilde{\omega}-m\tilde{\Omega}_{H}}{\tilde{\omega}}\right)|{\cal T}|^2.
\end{eqnarray}
For frequencies in the interval $0 < \tilde{\omega} <m\tilde{\Omega}_{H}$ the reflectance is always larger than unit, which implies in the superresonance phenomenon (analog to the superradiance in black hole physics)~\cite{Basak:2002aw,SBP}.
Here $m$ is the azimuthal mode number and $\Omega_{H}=Bc/A^2$ is the angular velocity of the usual Kerr-like acoustic black hole. Notice from
Eq. (\ref{refle}) that the frequency $\tilde{\omega}$ and the modified angular velocity $\tilde{\Omega}_{H}$ depends on the Lorentz violating parameter $\tilde{\beta}_{-} = 1-\beta$. This means that for $-1<\beta<1$ we have an increasing ($-1<\beta<0$) or decreasing ($0<\beta<1$) in the frequencies and a larger or smaller
spectrum of particles wave function can be scattered with increased amplitude.  For the previously assumed reasonable Lorentz-violating parameter $\beta\sim10^{-8}$ one expects a small effect. Alternatively, the acoustic Kerr-like black hole possesses a
wider or narrower rate of loss of mass (energy).
Thus, we show that the presence of the Lorentz violating parameter
modifies the quantity of removed energy of the
acoustic black hole and that is either possible to accentuate
or attenuate the amplification of the removed energy
of the acoustic black hole.
The effect of superresonance can be eliminated when $\tilde{\beta}_{+}=\frac{m\Omega_{H}}{\omega}$, or $\beta=\frac{m\Omega_{H}}{\omega}-1$. In this case, the reflection coefficient is equal to unity. 
A similar result was obtained in~\cite{GAM}, where the effect of superradiance in acoustic black hole was studied in the presence of disclination and
a correction dependent on the disclination in terms of the angular velocity  was obtained.

\subsection{The case $\beta=0$ and $\alpha\neq0$}
In the present subsection we repeat the previous analysis for $\beta=0$ and $\alpha\neq0$. As in the earlier case we take the acoustic line element with Lorentz symmetry breaking obtained in  \cite{ABP}  in the non-relativistic limit, up to first order in $\alpha$, given by
\begin{eqnarray}
ds^2=-\tilde{\alpha}\left[1-\frac{(v_{r}^2+v_{\phi}^2)}{\tilde{\alpha}c^2}\right]d\tau^2
+\tilde{\alpha}^{-1}\left(1-\frac{v_{r}^2}{\tilde{\alpha}c^2}\right)^{-1}dr^2
-\frac{2v_{\phi}}{c}rd\varphi d\tau+\left(1-2\alpha v\right)r^2d\varphi^2,
\end{eqnarray}
with, $2\alpha v=-2\alpha(v_{r}+v_{\phi})$. Thus, in the new coordinates the metric becomes
\begin{eqnarray}
\label{am}
ds^2=-\tilde{\alpha}\left(1-\frac{r_{e}^2}{r^2}\right)d\tau^2
+\tilde{\alpha}^{-1}\left(1-\frac{r_{h}^2}{r^2}\right)^{-1}dr^2
-\frac{2B}{cr}rd\varphi d\tau+\left[1+\frac{2\alpha(\tilde{\alpha}^{1/2}cr_{h}+B)}{r}\right]r^2d\varphi^2,
\end{eqnarray}
where $\tilde{\alpha}=1+\alpha$. 
The radius of the ergosphere $(r_{e})$ and the horizon $(r_{h})$ are given by
\begin{eqnarray}
r_{e}=\frac{1}{\tilde{\alpha}^{1/2}}\frac{\sqrt{{A^2+B^2}}}{c}, \quad r_{h}=\frac{|A|}{\tilde{\alpha}^{1/2}c}.
\end{eqnarray}
The components of the metric are
\begin{eqnarray}
g_{\mu\nu}=\left[\begin{array}{clcl}
-{\tilde\alpha}\left[1-\frac{r_{e}^2}{r^2}\right] &\quad\quad\quad 0& -\frac{B}{cr}\\
0 & \frac{1}{{\tilde\alpha}}\left(1-\frac{r_{h}^2}{r^2} \right)^{-1}& 0\\
-\frac{B}{cr} &\quad\quad\quad 0 & \left[1+\frac{2\alpha(\tilde{\alpha}^{1/2}cr_{h}+B)}{r}\right]
\end{array}\right],
\end{eqnarray}
and its inverse $g^{\mu\nu}$ reads
\begin{eqnarray}
\label{mtinva}
g^{\mu\nu}=\left[\begin{array}{clcl}
-\frac{\Theta(r)}{{\cal D}(r)} &\quad\quad 0& -\frac{B\Theta(r)\eta(r)}{cr{\cal D}(r)}\\
0 & \tilde{\alpha}\left(1-\frac{r_{h}^2}{r^2} \right)& 0\\
-\frac{B\Theta(r)\eta(r)}{cr{\cal D}(r)} &\quad\quad 0 & \tilde{\alpha}\left(1-\frac{r_{e}^2}{r^2}\right)\frac{\Theta(r)\eta(r)}{{\cal D}(r)}
\end{array}\right],
\end{eqnarray}
where $\Theta(r)=\left[1-\frac{r_{h}^2}{r^2}\right]^{-1}$, $\eta(r)=\left[1+\frac{2\alpha(\tilde{\alpha}^{1/2}cr_{h}+B)}{r}\right]^{-1}$ and ${\cal D}(r)=\frac{B^2\Theta(r)\eta(r)}{c^2r^2}+\tilde{\alpha}
[1-\frac{B^2\Theta(r)}{\tilde{\alpha}c^2r^2}]$. 

Now, the radial function $R(r)$, as in the previous case, satisfies the linear second order differential equation
\begin{eqnarray}
\label{EKG}
\left[\omega^2-\frac{2Bm\omega\eta(r)}{cr^2}
-\frac{\tilde{\alpha}m^2\eta(r)}{r^2}\left(1-\frac{r_{e}^2}{r^2}\right)\right]\frac{R(r)}{{\cal D}(r)}
+\frac{1}{r}\left(1-\frac{r_{h}^2}{r^2}\right)\frac{d}{dr}\left[r\tilde{\alpha}\left(1-\frac{r_{h}^2}{r^2}\right)
\frac{d}{dr}\right]R(r)=0.
\end{eqnarray}
Again, we introduce the tortoise coordinate $r^{\ast}$ through the equation
\begin{eqnarray}
\frac{d}{dr^{\ast}}=\left(1-\frac{r_{h}^2}{r^2}\right)\frac{d}{dr}.
\end{eqnarray}
Now, after introducing a new radial function, $G(r^{\ast})=r^{1/2}R(r)$, the modified radial equation (\ref{EKG}),
in the asymptotic region ($r^{\ast}\rightarrow\infty$), can be approximately written as
\begin{eqnarray}
\label{EGA}
\frac{d^2G(r^{\ast})}{dr^{\ast2}}+\tilde{\omega}^2G(r^{\ast})=0, \quad \tilde{\omega}^2=\frac{\omega^2}{\tilde{\alpha}^2}.
\end{eqnarray}
The solution of equation (\ref{EGA}) reads
\begin{eqnarray}
\label{slu}
G(r^{\ast})=e^{i\tilde{\omega}r^{\ast}}+{\cal R}e^{-i\tilde{\omega}r^{\ast}}\equiv G_{A}(r^{\ast}).
\end{eqnarray}
We are now able to compute the Wronskian of the solutions (\ref{slu}) to obtain
\begin{eqnarray}
\label{w+}
{\cal W}(+\infty)=-2i\tilde{\omega}(1-|{\cal R}|^2).
\end{eqnarray}
On the other hand, in the near horizon region ($r^{\ast}\rightarrow-\infty$), we have
\begin{eqnarray}
\frac{d^2G(r^{\ast})}{dr^{\ast2}}+\left(\tilde{\omega}-m\Omega_{H}\right)^2G(r^{\ast})=0,
\end{eqnarray}
where $\Omega_{H}=Bc/A^2$. In this region the solution reads as follows
\begin{eqnarray}
G(r^{\ast})={\cal T}e^{i\left(\tilde{\omega}-m\Omega_{H}\right)r^{\ast}}\equiv G_{H}(r^{\ast}),
\end{eqnarray}
and the Wronskian of this solution is
\begin{eqnarray}
\label{w-}
{\cal W}(-\infty)=-2i\left(\tilde{\omega}-m\Omega_{H}\right)|{\cal T}|^2,
\end{eqnarray}
so that, from (\ref{w+}) and (\ref{w-}), we obtain the reflection coefficient
\begin{eqnarray}
|{\cal R}|^2=1-\left(\frac{\tilde{\omega}-m\Omega_{H}}{\tilde{\omega}}\right)|{\cal T}|^2.
\end{eqnarray}
Note that only the frequency $\tilde{\omega}$ is affected by the parameter that violates the Lorentz symmetry and for $\alpha>0$ the frequency interval is increased. For the previously assumed reasonable Lorentz-violating parameter $\alpha\sim10^{-8}$ one expects a small effect. The angular velocity is not modified.
As in the previous example, for the extreme case $\tilde{\alpha}=\frac{\omega}{m\Omega_{H}}$ or $\alpha=\frac{\omega}{m\Omega_{H}}-1$, so that the reflection coefficient is equal to unity and the effect of superresonance is eliminated.

\section{PERFECT VORTEX}\label{PV}
Perfect vortex is obtained in the regime $A = 0$ in (\ref{ELB}). In this case this spacetime (acoustic vortex) represents
a fluid with a non-radial flow~\cite{LS}. The radius of the ergosphere is given by
\begin{equation}
r_{e}=\sqrt{\tilde{\beta}_{-}}\frac{B}{c}.
\end{equation}
In the limit $r\rightarrow\infty$, we have that the modified radial equation is given by
\begin{eqnarray}
\label{EGPV}
\frac{d^2G(r)}{dr^{2}}+\tilde{\omega}^2G(r)=0, \quad \tilde{\omega}^2=\frac{\tilde{\beta}_{+}^2}{\tilde{\beta}_{-}}\omega^2,
\end{eqnarray}
whose solution is
\begin{eqnarray}
G(r)=e^{i\tilde{\omega}r}+{\cal R}e^{-i\tilde{\omega}r},
\end{eqnarray}
and near the ergosphere $(r\rightarrow r_{e})$, the equation reads as follows
\begin{eqnarray}
\frac{d^2G(r)}{dr^{2}}+\left[\left(\tilde{\omega}-m\tilde{\Omega}_{e}\right)^2-\frac{\tilde{\Omega}_{e}^2}{4}\left(4m^2-1\right)\right]G(r)=0,
\end{eqnarray}
where $\tilde{\Omega}_{e}=\frac{1}{r_{e}}=\frac{c}{B\sqrt{\tilde{\beta_{-}}}}$,
is the angular velocity of the ergosphere. In this region, we have the solution as follows
\begin{eqnarray}
G(r)={\cal T}\exp\left[-i\tilde{\omega}\sqrt{\left(1-\frac{\tilde{\omega}_{+}}{\tilde{\omega}}\right)
\left(1-\frac{\tilde{\omega}_{-}}{\tilde{\omega}}\right)}r\right],
\end{eqnarray}
where
\begin{eqnarray}
\tilde{\omega}_{\pm}=m\tilde{\Omega}_{e}\left[1\pm\sqrt{1-(4m^2)^{-1}}\right].
\end{eqnarray}
Thus, we obtain the reflection coefficient
\begin{eqnarray}
|{\cal R}|^2=1-\sqrt{\left(1-\frac{\tilde{\omega}_{+}}{\tilde{\omega}}\right)
\left(1-\frac{\tilde{\omega}_{-}}{\tilde{\omega}}\right)}|{\cal T}|^2.
\end{eqnarray}
Note that it is impossible to have, $|{\cal R}|^2>1$. Thus, we have the absence of superresonance phenomenon for the perfect vortex. However, for 
$\beta\rightarrow1$ we have an increasing in the frequency $\tilde\omega$ and in the angular velocity of the ergosphere of an acoustic vortex (with pure rotation without horizons).

Let us now consider the second case described by the metric (\ref{am}). So for the perfect vortex in this case, i.e.,  with $A=0$, the radius of the ergosphere is
\begin{eqnarray}
r_{e}=\frac{B}{\tilde{\alpha}^{1/2}c}.
\end{eqnarray}
Now the reflection coefficient is given by
\begin{eqnarray}
|{\cal R}|^2=1-\sqrt{\left(1-\frac{\tilde{\omega}_{+}}{\tilde{\omega}}\right)
\left(1-\frac{\tilde{\omega}_{-}}{\tilde{\omega}}\right)}|{\cal T}|^2, 
\quad\tilde{\omega}=\frac{\omega\sqrt{\sigma}}{\tilde{\alpha}}
\end{eqnarray}
where
\begin{eqnarray}
\tilde{\omega}_{\pm}=m\tilde{\Omega}_{e}\left[1\pm\sqrt{1-(4m^2)^{-1}}\right], 
\quad \sigma=1+2\alpha c
\end{eqnarray}
 and $\tilde{\Omega}_{e}=\frac{c}{\sqrt{\sigma}B}$,
is the angular velocity of the ergosphere.

\section{Conclusions}
\label{conclu}

In this paper we have considered the implications of the metric obtained in \cite{ABP} for extended Abelian Higgs model with a Lorentz-violating term. As first shown in \cite{ABP}, one of the consequences is that the acoustic Hawking temperature is changed such that it depends on the group speed which means that, analogously to the gravitational case \cite{syb,adam}, the Hawking temperature is {\it not} universal for all species of particles. It depends on the maximal attainable velocity of this species. In the context of gravitational black holes this has been previously studied and appointed
as a sign of possibly violation of the second law of the thermodynamics.  Furthermore, the acoustic black hole metric in our model can be identified with 
an acoustic Kerr-like black hole. In the present analysis we explicitly have shown that the Lorentz violating term affects the rate of loss of mass (energy) of the black hole. We also have shown that for suitable values of
the Lorentz violating parameter a wider or lower spectrum of particle wave function can be scattered with increased amplitude by the acoustic black hole. This
increases the superressonance phenomenon previously studied in \cite{Basak:2002aw}. Thus, we show that the presence of the Lorentz violating parameter
modifies the quantity of removed energy of the acoustic black hole and that is either possible to accentuate or attenuate the amplification of the removed energy of the acoustic black hole,  though for the previously assumed reasonable Lorentz-violating parameter $\alpha,\beta\sim10^{-8}$ one expects a small effect.
In the extreme case, the effect of superresonance can be eliminated when $\tilde{\beta}_{+}=\frac{m\Omega_{H}}{\omega}$, or $\beta=\frac{m\Omega_{H}}{\omega}-1$. In this case, the reflection coefficient is equal to unity. The Abelian Higgs model is good to describe high energy physics
and extended Abelian Higgs model can also describe Lorentz symmetry violation in particle physics in high energy. Thus our results show that in addition to the expected gravitational mini black holes formed in high energy experiments one can also expect the formation of acoustic black holes together.

\acknowledgments

We would like to thank CNPq, CAPES, PNPD/PROCAD -
CAPES for partial financial support.

\end{document}